\documentclass[11pt,twoside]{article}
\usepackage{asp2010}

\resetcounters

\begin{document}

\title{Constraining the axion mass through the asteroseismology of the 
ZZ Ceti star G117$-$B15A} 

\author{A. H. C\'orsico$^1$, 
        L. G. Althaus$^1$, 
        A. D. Romero$^1$,
        M. M. Miller Bertolami$^1$, 
        E. Garc\'\i a--Berro$^{2,3}$, and 
        J. Isern$^{3,4}$
\affil{$^1$Facultad de Ciencias Astron\'omicas y Geof\'isicas, 
           Universidad Nacional de La Plata, 
           Paseo del Bosque s/n, 
           (1900) La Plata, Argentina}
\affil{$^2$Departament de F\'\i sica Aplicada, 
           Universitat Polit\`ecnica de Catalunya,
           c/Esteve Terrades 5, 
           08860 Castelldefels, 
           Spain}
\affil{$^3$Institute for Space Studies of Catalonia, IEEC,
           c/Gran Capit\`a 2-4, Edif. Nexus 104, 
           08034 Barcelona, 
           Spain}
\affil{$^4$Institut de Ci\`encies de l'Espai, CSIC, 
           Campus UAB, Facultat de Ci\`encies, Torre C-5, 
           08193 Bellaterra, 
           Spain}}

\begin{abstract}
We perform an asteroseismological study on the DAV star G117$-$B15A on
the basis of a modern set  of fully evolutionary DA white dwarf models
that have consistent  chemical profiles at the core  and the envelope.
We  found an  asteroseismological model  for G117$-$B15A  that closely
reproduces  its observed  pulsation  periods. Then,  we  use the  most
recently measured value of the  rate of period change for the dominant
mode of this pulsating star to impose a preliminary upper limit to the
mass of the axion.
\end{abstract}

\section{Introduction}

Pulsating DA (H-rich atmospheres) white dwarfs, also called ZZ Ceti or
DAV stars, are  the most numerous class of  degenerate pulsators, with
over 148 members known today.  They are characterized by multiperiodic
brightness  variations caused by  spheroidal, non-radial  $g$-modes of
low  degree with  periods between  70  and 1500~s  (Winget \&  Kepler,
Althaus et al.  2010a).  G117$-$B15A  is the most well-studied star of
this class of variables.  The rate  of change of its 215.2~s period is
very  small:  $\dot{\Pi}=(4.07\pm  0.61)\times 10^{-15}$  s/s  (Kepler
2009),  with  a  stability  comparable  to that  of  the  most  stable
millisecond pulsars.  The {\sl  axion} is a hypothetical particle that
appears as a consequence of the symmetry postulated by Peccei \& Quinn
(1977)  to solve  the  strong CP  (charge-parity)  problem in  quantum
chromodynamics.  Axions  are considered as candidates  for dark matter
of the Universe, and their contribution depends on their mass (Raffelt
2007).  Interestingly enough, axion emission is supposed to take place
in the cores of white dwarfs  (Isern et al.  1992, 2008). Since axions
can freely escape from such  stars, their existence would increase the
cooling rate and,  consequently, the rate of change  of the periods as
compared  with  the standard  one.   In this  work  we  present a  new
asteroseismological  model for  G117$-$B15A  and use  the more  recent
measurement of the rate of change  of the 215.2~s period to impose new
constraints on the mass of the axion.

\section{A new asteroseismological model for G117$-$B15A}

\begin{table*}
\centering
\caption{Characteristics of G117$-$B15A and of our seismological model.}
\begin{tabular}{lccc}
\hline
\hline
 Quantity             &    Koester  \& Holberg & Bergeron et al.   &   Our seismological  \\
                      &    (2001)              &     (2004)  &               model  \\          
\hline
$T_{\rm eff}$ [K]      &  $12\,010\pm 180$    & $11\,630\pm 200$ &   $11\,985 \pm 200$  \\
$M_*/M_{\odot}$        &  $0.55 \pm 0.10$     & $0.59 \pm 0.03$  &   $0.593 \pm 0.007$   \\
$\log g$              &  $7.94\pm 0.17 $     & $7.97\pm 0.05 $  &   $8.00\pm 0.09$   \\
$\log(R_*/R_{\odot})$  &   ---             &  ---             &   $-1.882\pm 0.029$  \\   
$\log(L_*/L_{\odot})$  &   ---             &  ---            &    $-2.497 \pm 0.030$  \\
$M_{\rm He}/M_*$       &   ---             &  ---            &   $2.39  \times 10^{-2}$  \\
$M_{\rm H}/M_*$        &   ---             &  ---            &   $(1.25\pm 0.7) \times 10^{-6}$   \\
$X_{\rm C},X_{\rm O}$ (center)   &  ---             &  ---            &   $0.28, 0.70$   \\
\hline
\end{tabular}
\label{table1}
\end{table*}

We  performed a  detailed asteroseismological  study of  the  DAV star
G117$-$B15A  using  a grid  of  evolutionary  models characterized  by
consistent  chemical profiles  for  both the  core  and envelope,  and
covering a wide  range of stellar masses, thicknesses  of the hydrogen
envelope  and  effective  temperatures.   This constitutes  the  first
asteroseismological application of the DA white-dwarf models presented
in Althaus et al.  (2010b).  These models were generated with the {\tt
LPCODE} evolutionary code from  the ZAMS through the thermally-pulsing
and mass-loss phases on the AGB and finally to the domain of planetary
nebulae and white dwarfs.  The effective temperature, the stellar mass
and the mass  of the H envelope  of our DA white dwarf  models vary in
the  ranges: $14\,000 \gtrsim  T_{\rm eff}  \gtrsim 9\,000$  K, $0.525
\lesssim  M_* \lesssim  0.877 M_{\odot}$,  $-9.4  \lesssim \log(M_{\rm
H}/M_*) \lesssim -3.6$, where the  ranges of the values of $M_{\rm H}$
are dependent on  $M_*$. For simplicity, the mass of  He has been kept
fixed at the value predicted by the evolutionary computations for each
sequence.

\begin{figure*} 
\begin{center}
\includegraphics[clip,width=13 cm]{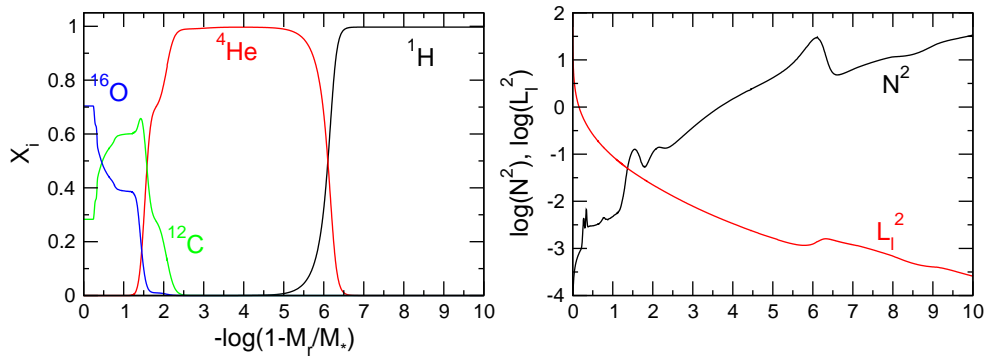} 
\caption{The  internal  chemical  stratification  (left panel)  and  a
  propagation diagram  (right panel) of  our asteroseismological model
  for G117$-$B15A.}
\label{figure1} 
\end{center}
\end{figure*}

We  searched for  a pulsation  model that  best matches  the pulsation
periods  of  G117$-$B15A.  To  this  end,  we  sought the  model  that
minimizes  a quality  function defined  simply as  the average  of the
absolute differences between theoretical and observed periods: $\Phi =
\Phi(M_*, M_{\rm H}, T_{\rm eff})= \frac{1}{N}\sum_{i=1}^N |\Pi_k^{\rm
th}-\Pi_i^{\rm obs}|$, where  $N= 3$ is the number  of the periods ---
215.20~s, 270.46~s and  304.05~s (Kepler et al. 1982)  --- observed in
G117$-$B15A.  The  theoretical periods were  assessed by means  of the
pulsation code  described in C\'orsico  \& Althaus (2006). We  found a
best-fit model  with the characteristics shown  in Table \ref{table1}.
The internal chemical stratification  and a propagation diagram (i.e.,
the spatial  run of the  logarithm of the  squared Brunt-V\"ais\"al\"a
and Lamb  frequencies) of this model are  shown in Fig.~\ref{figure1}.
Each  chemical  transition   region  produces  clear  and  distinctive
features   in  $N$,   which   are  eventually   responsible  for   the
mode-trapping properties of  the model. In the core  region, there are
several  peaks   at  $-\log(q)  \approx  0.4-0.5$   (where  $q  \equiv
1-M_r/M_*$) resulting  from steep variations in the  inner C-O profile
which are caused by the occurrence of extra mixing episodes beyond the
fully  convective core  during central  helium burning.   The extended
bump  in $N^2$ at  $-\log(q) \approx  1-2$ is  caused by  the chemical
transition of He,  C and O resulting from  nuclear processing in prior
AGB  and thermally-pulsing  AGB  stages. Finally,  there  is the  He/H
transition region at $-\log(q) \approx 6$, which is smoothly shaped by
the action of time-dependent element diffusion.

\section{A new upper limit for the axion mass}

In  order to compute  the effects  of axion  emission, we  adopted the
axion  emission  rates of  Nakagawa  et  al.  (1988).  Axion  emission
produces a supplementary energy loss rate to those normally considered
in the standard theory of  white dwarf evolution.  As a result, axions
accelerate the  cooling process of white dwarfs.  Such an acceleration
of  the  evolution  has   a  direct  consequence  on  the  pulsational
properties  of the star,  because it  produces a  larger value  of the
period derivative of the oscillation modes, $\dot{\Pi} \equiv d\Pi/dt$
(Isern  et  al.  1992).  Due  to  this  extra cooling  mechanism,  the
structure of the white dwarf itself is also affected but, fortunately,
in such  a way that  for a fixed  $T_{\rm eff}$ the  pulsation periods
$\Pi$ are largely  independent of the exact value  of the axion energy
losses (C\'orsico et al. 2001).

\begin{figure*} 
\begin{center}
\includegraphics[clip,width=11 cm]{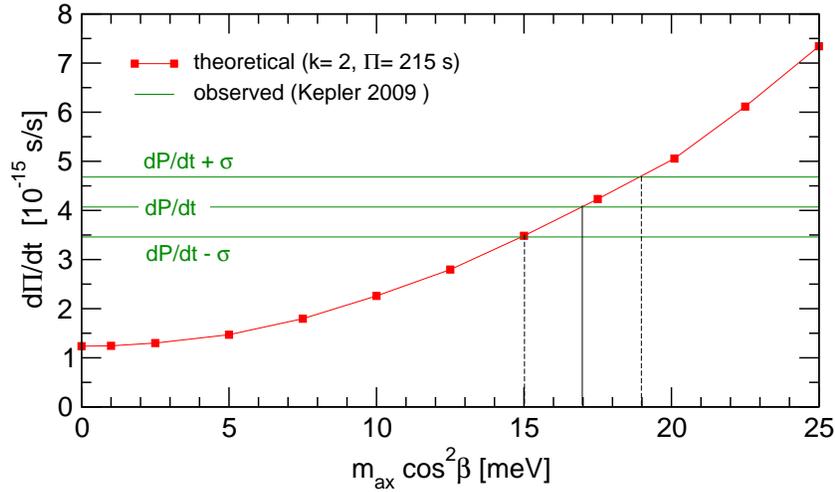} 
\caption{The temporal  derivative of  the period corresponding  to the
  $l$ = 1,  $k$ = 2 mode of our asteroseismological  model in terms of
  the  axion mass. The  horizontal green  lines indicate  the observed
  value with its uncertainties (Kepler 2009).}
\label{figure2} 
\end{center}
\end{figure*} 

We have computed the values  of $\dot{\Pi}$ for the modes with periods
$\sim 215$  s, $\sim 270$  s and $\sim  304$ s ($k=  2, 3, 4$)  of our
asteroseismological  model by adopting  different axion  emissions. As
expected, the  pulsation periods  themselves do not  vary appreciably,
thus  validating   our  asteroseismological  model   even  with  axion
emission, but  the rates of  period change monotonically  increase for
increasing  axion emission.   In  Fig.  \ref{figure2}  we display  the
theoretical  value of  $\dot{\Pi}$ corresponding  to the  period $\Pi=
215$~s.  Also  shown is the most  recent determination of  the rate of
period change for  the period at $\Pi= 215$  of G117$-$B15A, according
to Kepler (2009). Note that,  if we consider a standard deviation from
the  observational  value,  we  conclude  that  the  observations  are
compatible with an axion mass  lower than $\approx 19$~meV. This value
is about  5 times larger than  that found by C\'orsico  et al. (2001),
but  in  good   agreement  with  the  range  of   values  obtained  by
Bischoff-Kim et al.  (2008), which derived axion masses between 12 and
26.5~meV.

\section{Conclusions}

On the basis  of a new asteroseismological model  for G117$-$B15A, the
archetype  of   DAV  stars,  and   making  use  of  the   most  recent
determination of  the rate of period  change for the  dominant mode of
this star, we have derived a new upper limit of the mass of the (up to
now) elusive particle called  {\sl axion}. We emphasize, however, that
the  derived upper  limit for  the  axion mass  ($\approx 19$~meV)  is
preliminary,  since we have  not quantified  yet the  uncertainties of
white  dwarf  modeling  (core   overshooting,  rate  of  the  reaction
$^{12}$C$(\alpha,\gamma)^{16}$O) and the  uncertainties related to the
asteroseismological approach.

The main  conclusion of this preliminary  study is that,  if we assume
that  our  asteroseismological  model  is  a  good  representation  of
G117$-$B15A,  in   order  to  explain  the  high   observed  value  of
$\dot{\Pi}(215)$  it  is  necessary   to  invoke  some  extra  cooling
mechanism, being the axion emission a very plausible one.

\acknowledgements 

One  of us  (A.H.C.)   warmly thanks  Prof.   Hiromoto Shibahashi  for
support, that allowed him to attend the conference.

{}

\end{document}